\documentclass[preprint,prd,tightenlines,superscriptaddress]{revtex4-1}
\usepackage{graphicx} 
\usepackage{dcolumn}  
\usepackage{colordvi}
\usepackage{color}
\usepackage{epstopdf}
\usepackage{amssymb}
\usepackage{lineno}
\usepackage[percent]{overpic}
\usepackage{url}
\graphicspath{{ps}}
\usepackage{hyperref}
\usepackage{tabularx}
\usepackage{amsmath}
\usepackage{float} 
\usepackage{bm}
\usepackage{subfigure} 
\usepackage{comment} 
\usepackage[compat=1.0.0]{tikz-feynman}
\usepackage{tikz-feynman}
\usepackage{hyperref}
\usepackage{tasks}

\usepackage{placeins}

\usepackage[english]{babel}

\input belle2sym.tex

\begin{document}

\def\belletwo {\it {Belle II}}

\vspace*{-3\baselineskip}
\resizebox{!}{3cm}{\includegraphics{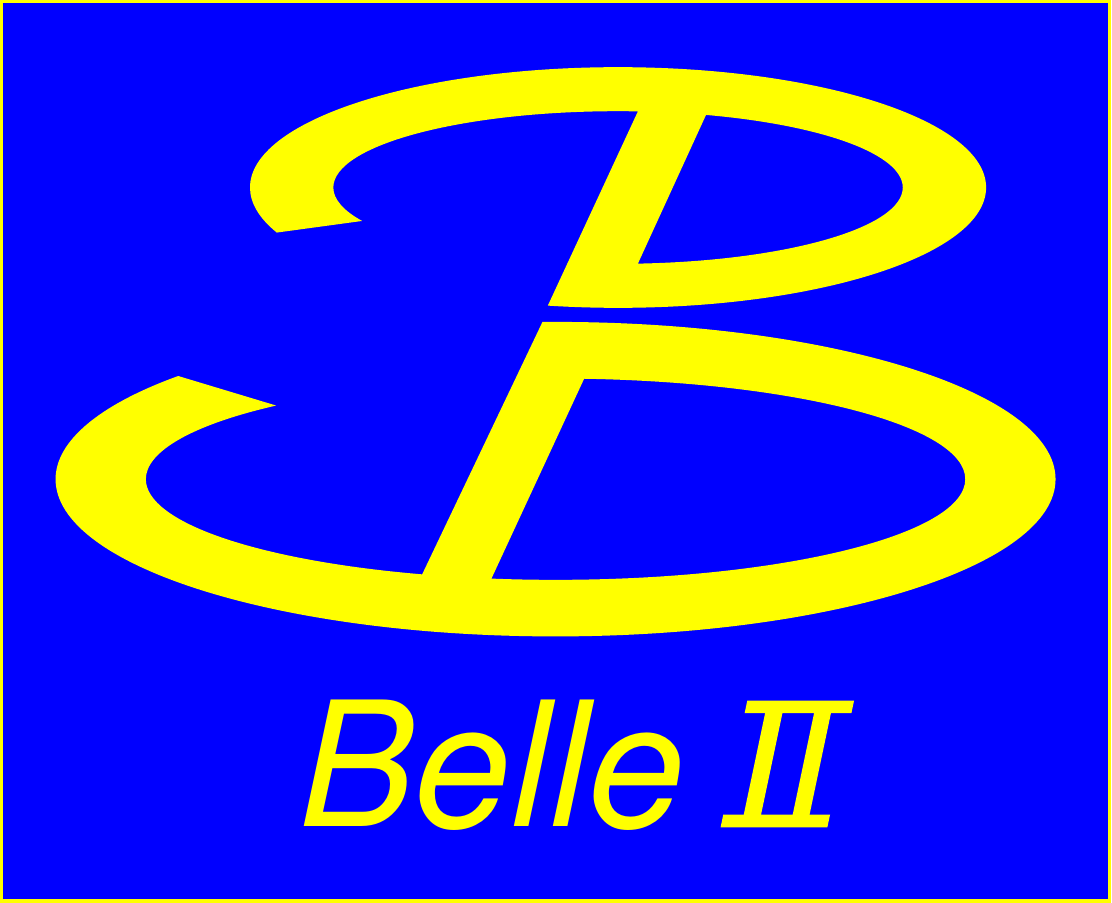}}

\vspace*{-5\baselineskip}
\begin{flushright}
BELLE2-CONF-PH-2022-009
\end{flushright}

\title { \quad\\[0.5cm] Measurement of the branching fraction for the decay $B \to K^{\ast}(892)\ell^+\ell^-$ at Belle II}
  \author{F.~Abudin{\'e}n}
  \author{I.~Adachi}
  \author{R.~Adak}
  \author{K.~Adamczyk}
  \author{L.~Aggarwal}
  \author{P.~Ahlburg}
  \author{H.~Ahmed}
  \author{J.~K.~Ahn}
  \author{H.~Aihara}
  \author{N.~Akopov}
% ...
\collaboration{Belle II Collaboration}

\begin{abstract}
\vspace{0.5cm}
We report a measurement of the branching fraction of $B \to K^{\ast}(892)\ell^+\ell^-$ decays, where $\ell^+\ell^- = \mu^+\mu^-$ or $e^+e^-$, using electron-positron collisions recorded at an energy at or near the $\Upsilon(4S)$ mass and corresponding to an integrated luminosity of $189$ fb$^{-1}$. The data was collected during 2019--2021 by the Belle II experiment at the SuperKEKB $e^{+}e^{-}$ asymmetric-energy collider. We reconstruct $K^{\ast}(892)$ candidates in the $K^+\pi^-$, $K_{S}^{0}\pi^+$, and $K^+\pi^0$ final states. The signal yields with statistical uncertainties are $22\pm 6$, $18 \pm 6$, and $38 \pm 9$ for the decays $B \to K^{\ast}(892)\mu^+\mu^-$, $B \to K^{\ast}(892)e^+e^-$, and $B \to K^{\ast}(892)\ell^+\ell^-$, respectively. We measure the branching fractions of these decays for the entire range of the dilepton mass, excluding the very low mass region to suppress the $B \to K^{\ast}(892)\gamma(\to e^+e^-)$ background and regions compatible with decays of charmonium resonances, to be
\begin{align*}
    {\cal B}(B \to K^{\ast}(892)\mu^+\mu^-) &= (1.19 \pm 0.31 ^{+0.08}_{-0.07}) \times 10^{-6},\\
    {\cal B}(B \to K^{\ast}(892)e^+e^-) &= (1.42 \pm 0.48 \pm 0.09)\times 10^{-6}, \\
    {\cal B}(B \to K^{\ast}(892)\ell^+\ell^-) &= (1.25 \pm 0.30  ^{+0.08}_{-0.07}) \times 10^{-6},
\end{align*}
where the first and second uncertainties are statistical and systematic, respectively. These results, limited by sample size, are the first measurements of $B \to K^{\ast}(892)\ell^+\ell^-$ branching fractions from the Belle II experiment. 

\end{abstract}


\maketitle

{\renewcommand{\thefootnote}{\fnsymbol{footnote}}}
\setcounter{footnote}{0}

\pagebreak

\section{Introduction}
The rare decays $B \to K^{\ast}(892)\ell^+\ell^-$ involve a $b \to s$ quark transition and are mediated by a flavor-changing neutral current. In the Standard Model (SM) they are forbidden at tree level and proceed through electroweak penguin or box amplitudes. As a consequence, these decays are highly suppressed and are sensitive to non-SM physics effects, which can enhance or suppress the amplitude of the decay or modify the angular distribution of the final-state particles. LHCb has made the most precise branching fraction measurement of $B^0 \to K^{\ast}(892)^{0}(K^+\pi^-)\mu^+\mu^-$ $i.e.,$ $(9.04^{+0.16}_{-0.15} \pm 0.62) \times 10^{-7}$~\cite{LHCb_BR}. The observation of these decays is the first step toward the measurement of $R_{K^{\ast}}$ in Belle~II. The lepton-flavor universality ratio is defined as the branching-fraction ratio of the muon to electron channel
	\begin{equation}
	R_{K^{*}}=\frac{\mathcal{B}(B \rightarrow K^{\ast}(892) \mu^+ \mu^-)}{\mathcal{B}(B \rightarrow K^{\ast}(892) e^+e^-)}.
	\label{rkstar_equation}
	\end{equation}
It is a sensitive probe for non-SM physics~\cite{sensitive1, sensitive2}. This $R_{K^{*}}$ observable is theoretically clean as the SM prediction is unity with a small theoretical uncertainty~\cite{sensitive1, rk_theory2}. On the other hand, measurements of $R_{K^{*}}$ for $q^2 \in [0.045-1.1]$ and $q^2 \in [1.1-6.0]$ GeV$^2/c^4$ differ by $2.1-2.3\sigma$ and $2.4-2.5\sigma$ from SM expectations~\cite{RKstar_LHCb}, which has attracted a lot of attention. While the precise measurement of $R_{K^{\ast}}$ constitutes an important goal of Belle II, at present we show the first observation of the $B \to K^{\ast}(892)\ell^+\ell^-$ decay and a measurement of its branching fraction using early electron-positron ($e^{+}e^{-}$) collision data. With a larger data sample we will measure the branching fraction more precisely as well as $R_{K^{\ast}}$ to check the consistency with the SM. The decay modes considered for this analysis are $B^0 \to K^{\ast 0}(892)(K^+\pi^-)\ell^+\ell^-$, $B^+ \to K^{\ast +}(892)(K_{S}^{0}\pi^+, K^+\pi^0)\ell^+\ell^-$, where $\ell^+\ell^- = \mu^+\mu^-$ or $e^+e^-$. The inclusion of the charge-conjugate decay mode is implied. From now on $K^{\ast}$ will be used as a shorthand for $K^{\ast}(892)^{0}$ and $K^{\ast}(892)^{+}$.

\section{The Belle II Detector and Data set}
Belle II is a large-solid-angle magnetic spectrometer designed to study products of energy-asymmetric $e^+e^-$ collisions at a center-of-mass energy corresponding to the mass of the $\Upsilon(4S)$ resonance. The detector is located at the interaction point of the SuperKEKB accelerator~\cite{superkekb}. The detector's components are arranged in a cylindrical geometry around the beam pipe. The innermost region of the detector comprises two subdetectors, namely two layers of pixel detectors and four layers of double-sided silicon strip detectors. The combination of pixel detector and silicon strip detector constitutes the inner tracking system. The measurement of charge and momentum of charged particles is provided by a 56-layer central drift chamber, which also provides particle identification (PID) information by measuring specific ionization. A Cherenkov-light angle and time-of-propagation detector situated in the barrel region and a proximity-focusing aerogel ring-imaging Cherenkov counter placed in the forward region together constitute the core of the PID system. An electromagnetic calorimeter (ECL) consisting of CsI(Tl) crystals measures the energy of photons and provides electron identification. These subdetectors are located inside a superconducting solenoid coil with a 1.5\,T magnetic field. The return yoke of the magnet is instrumented with plastic scintillators and resistive plate chambers to identify $K^{0}_{\rm L}$ mesons and muons, forming the KLM subdetector. Further details about the detector can be found in Ref.~\cite{belle2tdr}. 

The data sample used in this analysis was collected by Belle II in the period of 2019--2021 at the $\Upsilon(4S)$ resonance. The integrated luminosity is $189$ fb$^{-1}$, which is equivalent to $197 \times 10^6$ $B \overline{B}$ events. To study the properties of signal events, optimize selection criteria, and determine detection efficiencies, we use ten million simulated $B\overline{B}$ events in which one of the $B$ decays to the channel of interest and the other $B$ decays generically. This sample is referred to as signal Monte Carlo (MC). The events are generated using the EvtGen~\cite{{EVETGEN}} package. Geant4~\cite{GEANT4} is used to simulate the detector response. In addition, inclusive $B\overline{B}$ and  $q\overline{q}$ continuum ($q$ denotes $u, d, s, \text{ and } c$ quark) MC samples, equivalent to an integrated luminosity of $1~\mathrm{ab}^{-1}$, are used for background classification. The Belle II analysis software framework~\cite{basf2} is used to process the simulated and collision data.

\section{Event selection and Reconstruction}
Muons, electrons, and charged pions and kaons are reconstructed from charged particles originating near the $e^+e^-$ collision point, with a distance of closest approach in the plane transverse to the beam axis ($xy$ plane) $|d_0| < 2.0$ cm and along the beam axis ($z$ direction) $|d_z| < 4.0$ cm. A charged particle is identified as a $K^{\pm}$ or $\pi^{\pm}$ using ${\mathcal{P}(K/\pi)=\dfrac{L_{K}}{L_{K}+L_{\pi}}}$, where ${L_{K}}$ and ${L_{\pi}}$ are the likelihoods for the observed track to be consistent with a kaon or a pion, calculated by using information from all PID subdetectors. We require ${\mathcal{P}(K/\pi)> 0.6}$ to select ${K^{+}}$ and ${\mathcal{P}(\pi/K)> 0.6}$ to select ${\pi^{+}}$ candidates. The kaon (pion) selection efficiency is $86\%$ $(91\%)$ and has a misidentification rate of $7\%~(8\%)$ for pions (kaons). The $\ell^{\pm}$ candidates are identified using ${\mathcal{P}(\ell)=\dfrac{L_{\ell}}{L_{e}+L_{\mu}+L_{\pi} + L_{K}+ L_{p}+L_{d}}}$, where $L_{e}$, $L_{\mu}$, $L_{\pi}$, $L_{K}$, $L_{p}$, and $L_{d}$ are the likelihoods of a track being an electron, muon, pion, kaon, proton, and deuteron, respectively. We select the muon candidates having ${\mathcal{P}(\mu)}>0.9$ and a minimum momentum $p(\mu)$ of $0.8$~GeV$/c$ to ensure the particle reaches the KLM subdetector, corresponding to an efficiency of $87\%$ with $7\%$ pion misidentification rate. Similarly, electron candidates are required to satisfy ${\mathcal{P}(e)}>0.9$ and $p(e)>0.4$~GeV$/c$. This selection has an efficiency of $94\%$ with a $2\%$ pion misidentification rate. For the electron selection, information from all subdetectors is used except the time-of-propagation subdetector to have better electron-pion separation. To recover energy loss due to possible bremsstrahlung, we search for photons inside multiple cones centered around the electron momentum direction; we also require the cluster energy of the selected photons to be greater than $0.075$, $0.05$, and $0.1$~GeV in the forward endcap, barrel, and backward endcap ECL region, respectively. These requirements suppress low-energy photons resulting from particle interactions with detector material or the beam pipe. The $K^0_{\rm S}$ candidates are reconstructed from pairs of oppositely-charged particles, assumed to be pions, and are kinematically fit assuming they originate from a common vertex. In addition, $K_{S}^{0}$ are further selected using a boosted decision tree classifier~\cite{r13} that exploits momentum-dependent criteria on the $K_{S}^{0}$ flight length in the transverse plane, the azimuthal angle between the momentum vector and the vector between the interaction point and the decay vertex of the $K_{S}^{0}$ candidate, and the difference between $d_{z}$ of the two tracks. The invariant mass for $K^0_{\rm S}$ candidates is required to be in the range $[0.4876 - 0.5076]$~${\rm GeV}\!/c^2$, approximately $\pm 3\sigma$ about the $K^0_{\rm S}$ known mass. The $\pi^{0}$ candidates are reconstructed from pairs of photons each having an energy greater than 80, 30, or 60~MeV depending on whether the photon is detected in the forward, barrel, or backward region of the ECL, respectively. The higher energy requirements in the endcaps suppress low-energy beam background photons. Additional requirements such as ECL-cluster polar angle, number of crystals associated with the cluster, azimuthal angle difference between the two final state photons, and angle between the photons are imposed to suppress random $\gamma\gamma$ pairs. The invariant mass for $\pi^{0}$ candidates is required to be in the range $[0.1215 - 0.1415]$~${\rm GeV}\!/c^2$, which is $\pm 3\sigma$ around the $\pi^{0}$ known mass. 

The $K^{*}$ candidates are reconstructed by combining a kaon ($K^{+}$ or $K^0_{\rm S}$) with a pion ($\pi^{-}$ or $\pi^{0}$). We retain $K^{*}$ candidates inside the invariant-mass window $[0.796, 0.996] \,{\rm GeV}\!/c^2$, which corresponds to around four times the natural width of the $K^{*}$ meson. A $K^{*}$ is combined with two oppositely charged leptons to form a $B$ meson. To distinguish signal from background events, two kinematic variables are used, the beam-energy-constrained mass $M_{\rm bc} = \sqrt{E^{*2}_{\rm beam} - p^{*2}_{B}}$ and the energy difference $\Delta E = E^{*}_{B} - E^{*}_{\rm beam}$. Here, $E^{*}_{\rm beam}$ is the beam energy, and $p^{*}_{B}$ and $E^{*}_{B}$ are the momentum and energy of the $B$ meson, respectively. These quantities are calculated in the $e^+e^-$ center-of-mass frame. The constraints on these variables are $5.2 < M_{\rm bc} < 5.29$~$\rm GeV\!/c^{2}$ and $-0.15< \rm \Delta E < 0.1$~$\rm GeV$. For signal events, the $M_{\rm bc}$ distribution peaks at the nominal $B$ meson mass and the $\Delta E$ distribution peaks at zero.  

\section{Background suppression}
The major sources of backgrounds are from charmonium resonances, continuum, inclusive $B\overline{B}$, and events that mimic the signal decay (peaking background). The background from $B \to J/\psi(\ell^+\ell^-) K^{\ast}$ and $B \to \psi(2S)(\ell^+\ell^-) K^{\ast}$ decays is suppressed by the dilepton mass ($M(\ell^+\ell^-$)) vetoes
\begin{center}
\begin{tabular}{rl}
     $J/\psi$:& $M(\mu^+\mu^-) \notin [2.946,3.176]$ GeV$/c^2$, $M(e^+e^-) \notin [2.846,3.176]$ GeV$/c^2$, and\\
     $\psi(2S)$:& $M(\mu^+\mu^-) \notin [3.539, 3.719]$ GeV$/c^2$, $M(e^+e^-) \notin [3.439,3.719]$ GeV$/c^2$.
     \end{tabular}
\end{center}
The vetoed $B \to J/\psi(\ell^+\ell^-) K^{\ast}$ events are used as control channel. The background from $B \to K^{\ast}\gamma$ decays to the $B \to K^{\ast}e^+e^-$ channel due to photon conversion is suppressed by requiring the dilepton mass to be $M(e^+e^-)>0.14$~GeV$/c^2$. Two major sources of background remain, continuum and inclusive $B\overline{B}$ events. For continuum background, light-quark pairs produce two back-to-back jets because they have a lower mass and hence higher momentum compared to $B$ mesons. The main sources of backgrounds from inclusive $B\overline{B}$ processes are from opposite-side or same-side semileptonic decay, hadronic $B$ decays where one or more particles are misidentified as leptons, and background from misreconstructed $K^{\ast}$'s due to misidentification and swapping of daughter particles. In opposite-side semileptonic decay both $B$ mesons decay semileptonically and the final state leptons are misreconstructed as signal, whereas in same-side semileptonic decay one $B$ meson decays semileptonically followed by semileptonic decay of a daughter $D$ meson, and the final state leptons are misreconstructed as signal. 

A multivariate discriminant based on a boosted decision tree~\cite{r13} is used to separate signal and background. It is trained separately for each mode using the following event shape, vertex quality, and kinematic variables, known to provide statistical discrimination between signal and background: 
\begin{itemize}
    \item ratio of the second to zeroth Fox-Wolfram moments~\cite{foxwolfram},
    \item cosine of the angle between the $B$ flight direction and the beam axis,
    \item cosine of the angle between the thrust axis of the $B$ candidate and that of the rest of the event in the CM frame~\cite{thrust},
    \item magnitude of the signal $B$ thrust,
    \item longitudinal separation between the signal $B$ decay vertex and that of the other $B$,
    \item separation between the two lepton tracks in the $z$-direction,
    \item signal $B$ vertex probability,
    \item sum of the energies of the tracks and clusters of the rest of the event, and
    \item momentum flow into concentric cones around the thrust axis of a reconstructed $B$ candidate~\cite{cleocone}.
\end{itemize}
For training and testing the MVA, we use two independent MC data sets, each having an equal number of background and correctly reconstructed signal events. For each mode, a selection is applied on the MVA output to maximize 
the figure of merit $\dfrac{S}{\sqrt{(S+B)}}$, where $S$ and $B$ are the expected number of signal and background events in the signal region determined from simulation. The MVA rejects approximately $98\%$ of the background, with an approximate signal loss ranging between $30\%$ and $35\%$ depending on the decay channel. 

For the dimuon decay modes, misreconstructed events for which $M_{\rm{bc}}$ and $\Delta E$ lie in the signal region are observed. These backgrounds are due to hadrons misidentified as muons and mistakenly associated with the signal decay. They are suppressed using vetoes. The veto windows are decided by changing particle mass hypotheses depending on the decay mode. For the $B^0 \to K^{\ast0} (K^+\pi^-)\mu^+\mu^-$ channel, the following event veto selections are applied to suppress peaking background: (a) a veto on the pion-muon invariant mass  $M(\pi^-\mu^{+}) \notin [3.06, 3.11]$~GeV$/c^2$ is applied to suppress $B^0 \to J/\psi(\mu^+\mu^-)K^{\ast0}(K^+\pi^-)$ decays where the $\pi^-$ is misidentified as a $\mu^-$, and the $\mu^-$ as a $\pi^-$; (b) a veto on the kaon-pion-muon invariant mass $M(K^+\pi^-\mu^{-}) \notin [1.86, 1.885]$~GeV$/c^2$ is applied to suppress $B^0 \to D^{-}(K^+\pi^-\pi^-)\pi^+$ decays where a $\pi^-$ from $D^-$ decay and the $\pi^+$ from $B$ decay are misidentified as muons; (c) a veto on the kaon-muon invariant mass $M(K^+\mu^{-}) \notin [1.82, 1.9]$~GeV$/c^2$ is applied where the $\pi^-$ from $\overline{D^0}\to K^+\pi^-$ decays is misidentified as a $\mu^-$. For the $B^+ \to K^{\ast +} (K^0_{\rm S}\pi^+)\mu^+\mu^-$ channel,  two vetoes are applied: (a) a veto on the pion-muon invariant mass $M(\pi^+\mu^{-}) \notin [3.085, 3.105]$~GeV$/c^2$ is applied to suppress  $B^+ \to J/\psi (\mu^+\mu^-)K^{\ast +}(K^0_{\rm{S}}\pi^+)$ decays where the $\pi^+$ candidate is misidentified as a $\mu^+$ and vice versa; (b) a veto in the kaon-pion-muon invariant mass, $M(K_{S}^{0}\pi^+\mu^{-}) \notin [1.857,1.87]$~GeV$/c^2$ is applied to suppress $B^+ \to \overline{D^{0}}(K^0_{\rm{S}}\pi^+\pi^-)\pi^+$ decays where a $\pi^-$ from $\overline{D^0}$ decay and the $\pi^+$ from $B$ decay are misidentified as muons. For the $B^+ \to K^{\ast +} (K^+\pi^0)\mu^+\mu^-$ channel, a veto on the kaon-pion-muon invariant mass $M(K^+\pi^0\mu^-) \notin[ 1.855,1.87]$~GeV$/c^2$ is applied to suppress $B^+ \to \overline{D^{0}}(K^+\pi^0\pi^-)\pi^+$ decays where a $\pi^-$ from $\overline{D^0}$ decay and a $\pi^+$ from $B$ decay are misidentified as muons. These vetoes suppress the peaking backgrounds with a loss of $0.5\%-6.0\%$ in signal efficiency, depending on the decay mode.

After applying all the selection criteria, the candidate multiplicity per event ranges between $1.04$ and $1.24$, it is higher for the $B^+ \to K^{\ast +}(K^+\pi^0)\ell^+\ell^-$ channel due to fake $\pi^0$'s. In case of multiple candidates, we retain the one having $|\Delta E|$ closest to zero. The efficiency to select the correctly reconstructed signal from an event with multiple reconstructed $B$ candidates varies from $50-85\%$ depending on the decay mode. We check in simulation that the bias introduced by this criterion is negligible in our sample.

\section{Signal yield extraction}
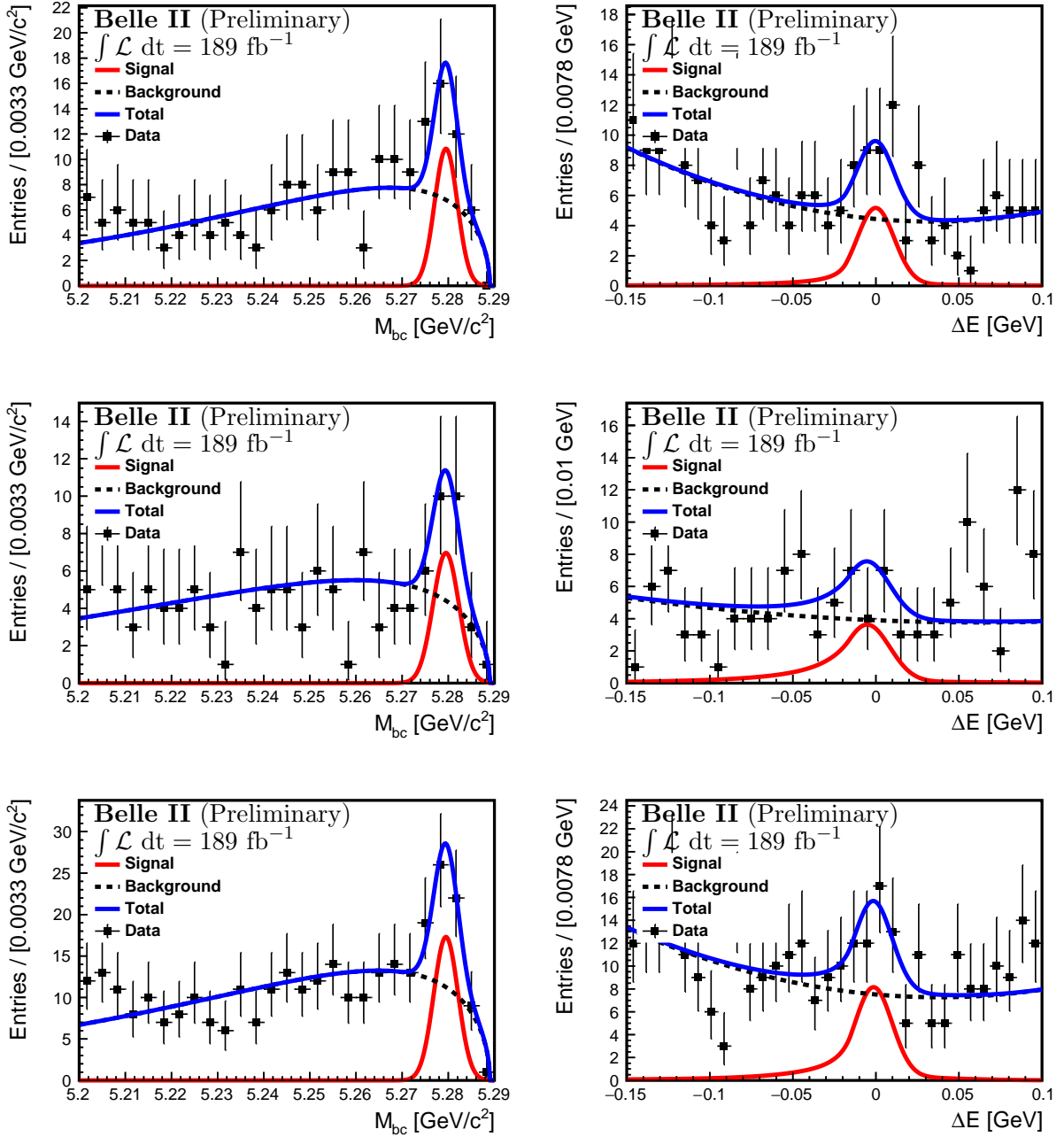
\begin{figure}  [H]
\begin{center}
\begin{overpic}[width=1\textwidth]{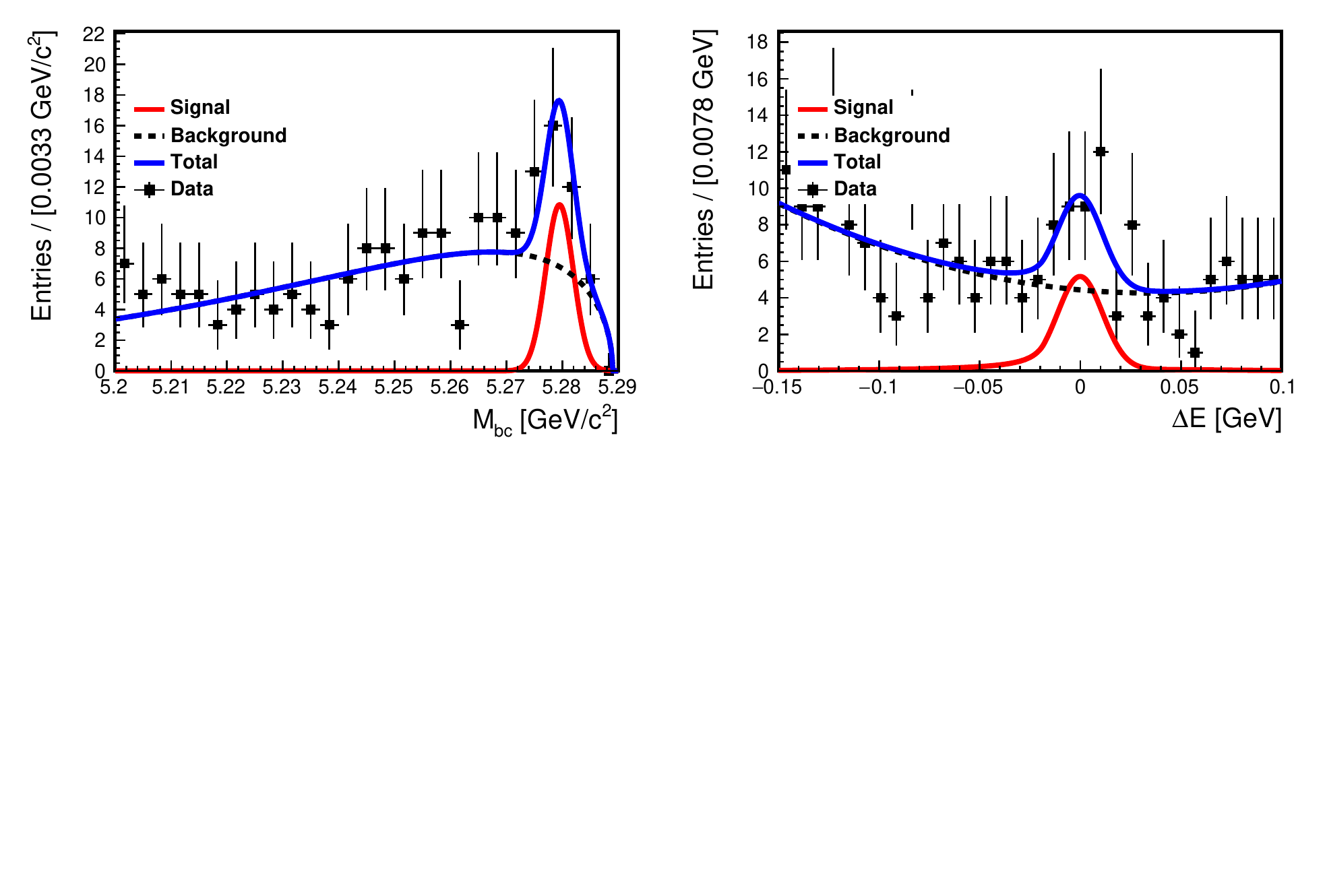}
 \put (10,63.2) {{\bf{Belle~II}} (Preliminary)}
 \put (10,60.7) {$\int \mathcal{L}~\rm{dt}=189~\rm{fb}^{-1}$}
 \put (60,63.2) {{\bf{Belle~II}} (Preliminary)}
 \put (60,60.7) {$\int \mathcal{L}~\rm{dt}=189~\rm{fb}^{-1}$}
\end{overpic}
\end{center}
\vspace{-6cm}
\begin{center}
\begin{overpic}[width=1\textwidth]{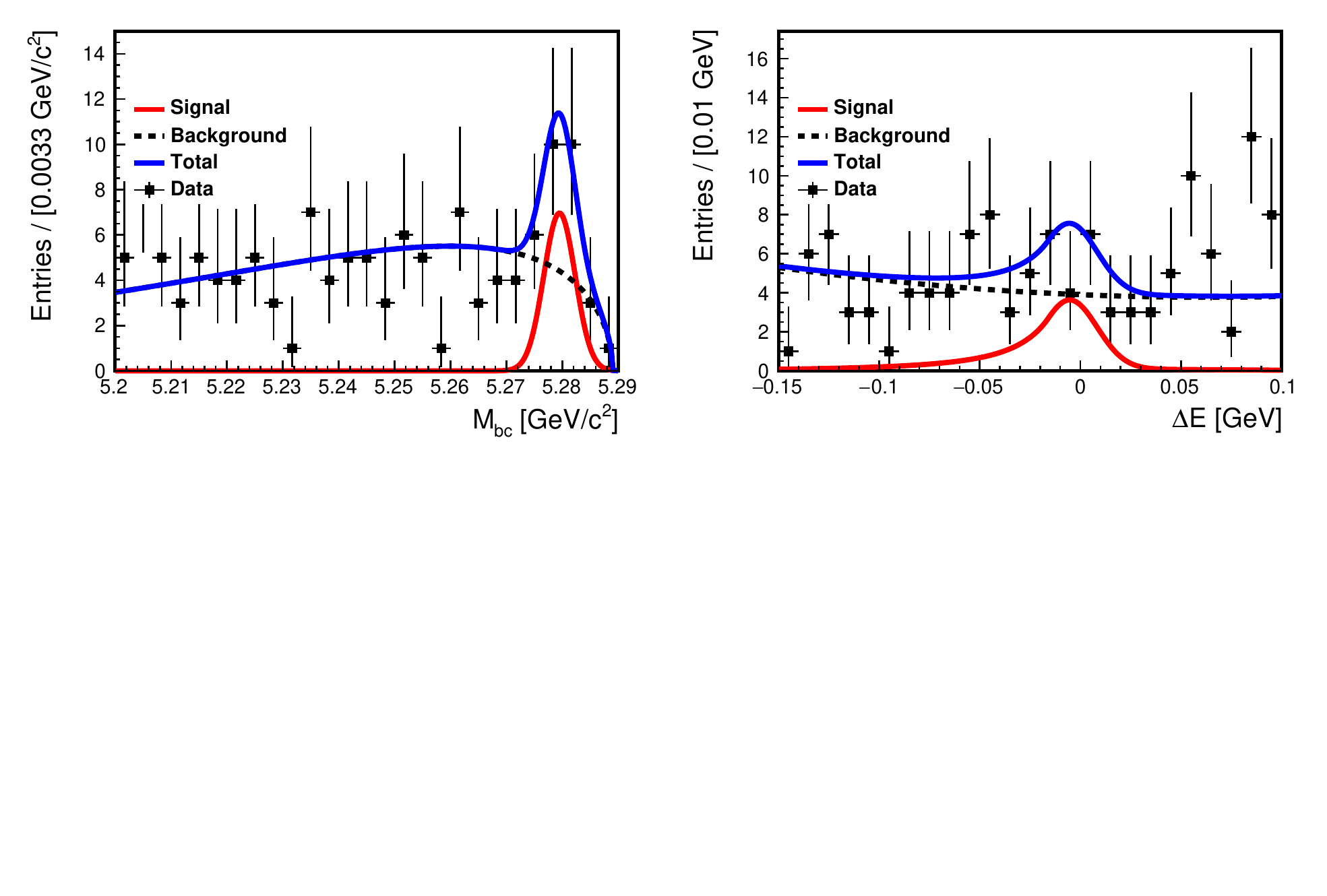}
 \put (10,63.2) {{\bf{Belle~II}} (Preliminary)}
 \put (10,60.7) {$\int \mathcal{L}~\rm{dt}=189~\rm{fb}^{-1}$}
 \put (60,63.2) {{\bf{Belle~II}} (Preliminary)}
 \put (60,60.7) {$\int \mathcal{L}~\rm{dt}=189~\rm{fb}^{-1}$}
\end{overpic}
\end{center}
\vspace{-6cm}
\begin{center}
\begin{overpic}[width=1\textwidth]{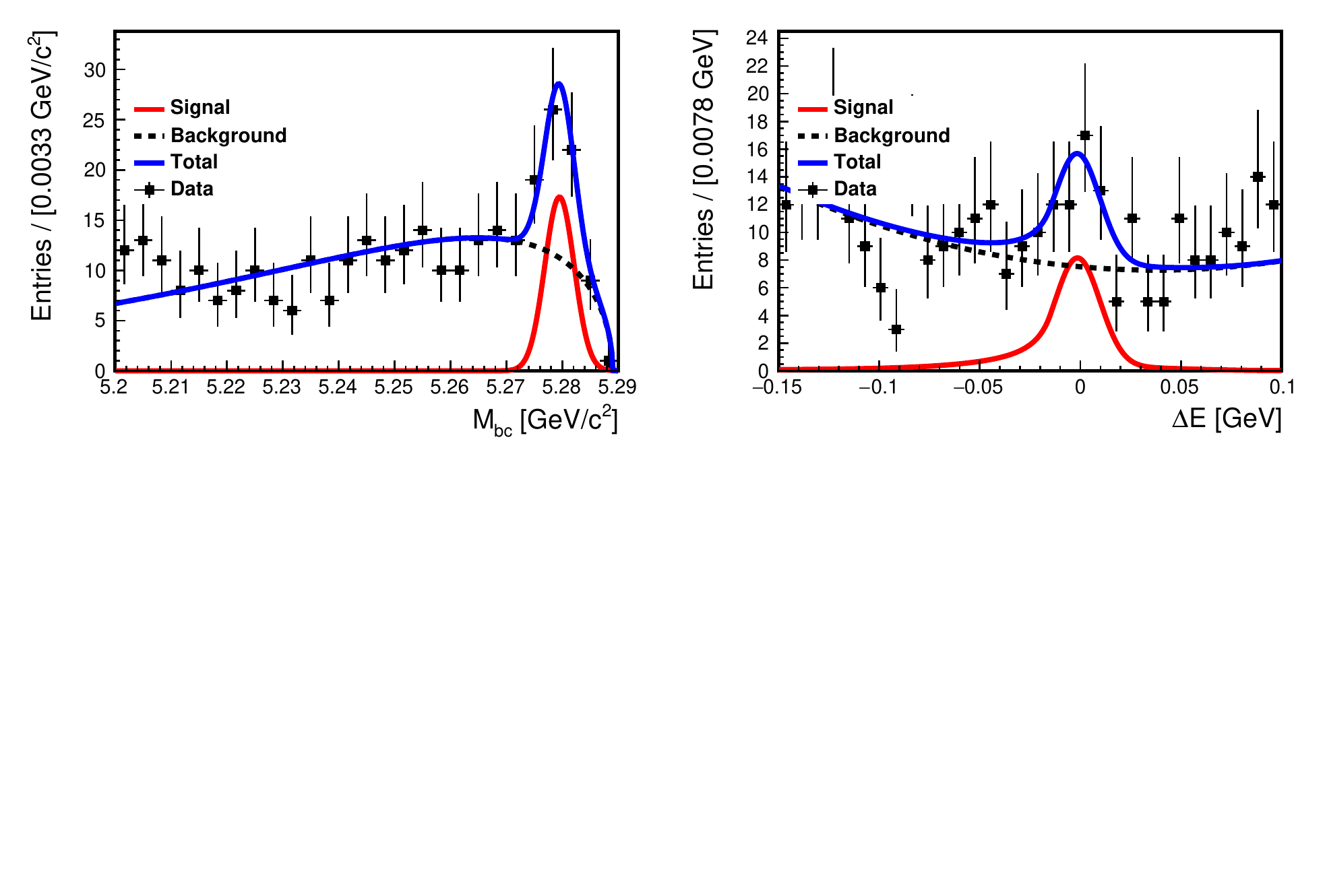}
 \put (10,63.2) {{\bf{Belle~II}} (Preliminary)}
 \put (10,60.7) {$\int \mathcal{L}~\rm{dt}=189~\rm{fb}^{-1}$}
 \put (60,63.2) {{\bf{Belle~II}} (Preliminary)}
 \put (60,60.7) {$\int \mathcal{L}~\rm{dt}=189~\rm{fb}^{-1}$}
\end{overpic}
\end{center}
\vspace{-6cm}
\caption{Distributions of $M_{\rm{bc}}$ (left) and $\Delta E$ (right) for $B \to K^{\ast}\mu^+\mu^-$ (top), $B \to K^{\ast}e^+e^-$ (middle), and $B \to K^{\ast}\ell^+\ell^-$ (bottom). Points with error bars are superimposed on the blue (solid) curve, which shows the total fit function, while red (solid) and black (dotted) lines represent the signal and background components, respectively. Candidates shown in the $\Delta E$ distributions are restricted to $M_{\rm{bc}} \in [5.27, 5.29]$~GeV$/c^{2}$ range and the $M_{\rm{bc}}$ distributions are restricted to $\Delta E \in [-0.05, 0.05]$~GeV.}
\label{Data_fit}
\end{figure}

The signal yields are extracted from two-dimensional extended maximum-likelihood fits to the unbinned $M_{\rm{bc}}$ and $\Delta E$ distributions, combining both charged and neutral $B$ samples. The probability density functions (PDFs) for correctly reconstructed signal events for $M_{\rm{bc}}$ and $\Delta E$ are modeled with a Gaussian and the sum of a Crystal ball~\cite{CB} and a Gaussian, respectively. The backgrounds are modeled with the ARGUS shape~\cite{argus} and a straight line for $M_{\rm{bc}}$ and $\Delta E$, respectively. The signal PDF parameters are fixed to those of the $B \to J/\psi(\ell^+\ell^-)K^{\ast}$ control channel and the background parameters are determined by the fit to data. The fit procedure is validated on simplified simulated experiments as well as on events reconstructed in the $B \to J/\psi(\ell^+\ell^-)K^{\ast}$ control channel. Signal-enhanced projection plots for $B \to K^{\ast}\mu^+\mu^-$, $B \to K^{\ast}e^+e^-$, and $B \to K^{\ast}\ell^+\ell^-$ are shown in Fig.~\ref{Data_fit}.

\section{Efficiency corrections and Systematic uncertainties}
\label{systematic}
The difference in efficiency between data and simulation due to the PID selection for charged hadrons is calculated using a $D^{\ast +} \to D^{0}(K^{-}\pi^{+})\pi^{+}$ control sample. The corrections are calculated as functions of momentum and cosine of the polar angle and ranges from $97\%-99\%$. We assign a systematic uncertainty of $0.4\%$ for kaon and $2.5\%$ for pion selection, respectively. The uncertainty due to lepton identification is investigated as a function of momentum and polar angle using $J/\psi \to \ell^+\ell^-$, $e^+e^- \to e^+e^-\ell^+\ell^-$, and $e^+e^- \to e^+e^-$ samples. The correction to the signal MC efficiency is in the range $91\%-98\%$ and the associated systematic uncertainties are $^{+1.9}_{-0.8}\%$ and $^{+0.9}_{-0.5}\%$ for muon and electron identification, respectively. An uncertainty of $2.0\%$ is assigned from comparing the reconstruction efficiency of $K^0_{\rm S}$ between data and simulation, using $D^{\ast +} \to D^0(K_{S}^{0}\pi^+\pi^-)\pi^+$ decays; the correction factor is compatible with one. The difference in reconstruction efficiency for $\pi^{0}$'s between data and simulation is studied by comparing the yield of $\eta \to \pi^{0}\pi^{0}\pi^{0}$ decays, and an uncertainty of $3.4\%$ is assigned for $\pi^{0}$ selection; the correction factor is compatible with one. We assign a systematic uncertainty of $0.3\%$ for each charged particle using an $e^+e^- \to \tau^+\tau^-$ sample, which accounts for the data-simulation discrepancies in the reconstruction of charged particles, and we linearly add this systematic uncertainty for each final-state charged particle. The uncertainty due to the requirements on MVA criteria to suppress background is studied in the $B \to J/\psi(\ell^+\ell^-)K^{\ast}$ control channel, and is $1.3\%-1.7\%$ depending on the decay mode, with a correction of $95\%-99\%$. The uncertainty due to the limited sample size of signal MC is less than $0.5\%$. The systematic due to signal cross feed (misreconstructed signal candidates) is found to be less than $1\%$ and is obtained by varying the fraction determined from signal MC within $\pm 50\%$, and modeling the PDF accordingly. The deviation of the signal yield from the nominal fit value is given as a systematic uncertainty. Shape parameters of the PDFs fixed in the nominal fit are varied by $\pm 1\sigma$ around their mean values, and the $0.5-1.0\%$ change in signal yield with respect to the nominal fit result is taken as a systematic uncertainty. The systematic uncertainty in the branching fractions of ${\cal B} (\Upsilon (4S) \to B^+B^-)$ and ${\cal B} (\Upsilon (4S) \to B^0\overline{B^0})$ is $1.2\%$~\cite{PDG}. A systematic uncertainty of $2.9\%$ is assigned to the estimate of the number of $B\overline{B}$ events. 
We summarize the systematic uncertainties in Table~\ref{table:Systematic}. The individual sources of uncertainties are assumed to be independent and the corresponding uncertainties are added in quadrature to determine the total uncertainty.

\begin{table}[htb]
\caption{ Relative systematic uncertainties (in \%) for $B \to K^{\ast}\ell\ell$.}
\vspace{-0.3cm}
\label{table:Systematic}
\begin{center}
\begin{tabular}{   l  c}
\hline Source &  Systematic ($\%$)  \\ \hline 
Kaon identification & $0.4$\\
Pion identification & $2.5$\\
Muon identification & $^{+1.9}_{-0.8}$\\
Electron identification & $^{+0.9}_{-0.5}$ \\
$K_{S}^{0}$ identification & $2.0$\\
$\pi^0$ identification & $3.4$\\
Tracking& $1.2-1.5$\\
MVA selection & $1.3 - 1.7$\\
Simulated sample size & $<0.5$\\
Signal cross feed & $<1\%$ \\
Signal PDF shape   & $0.5-1.0\%$\\
${\cal B} (\Upsilon (4S) \to B^+B^-)[({\cal B} (\Upsilon (4S) \to B^0\overline{B^0}))$ & $1.2$\\
Number of $B\overline{B}$ pairs& $2.9$\\ \hline
Total & $^{+6.7}_{-6.0}$\\ \hline
\end{tabular}
\end{center}
\end{table}

\section{Results and Summary}
We reconstruct $22 \pm 6$, $18 \pm 6$, and $38 \pm 9$ signal events for $B \to K^{\ast}\mu^+\mu^-$, $B \to K^{\ast}e^+e^-$, and $B \to K^{\ast}\ell^+\ell^-$ corresponding to $4.8\sigma$, $3.6\sigma$, and $5.9\sigma$, respectively, here $\sigma$ denotes the significance from a null yield and is defined as $\sigma = \sqrt{-2 \ln(\mathcal{L}_{0}/\mathcal{L})}$, where $\mathcal{L}_{0}$ is the likelihood with $N_{\rm{sig}}$ constrained to be zero and $\mathcal{L}$ is the maximum likelihood, using $189$~fb$^{-1}$ data collected in the 2019--2021 run period. Here, the uncertainties are statistical only. The branching fraction is calculated using the formula
\begin{center}
    ${\cal B}(B \to K^{\ast}\ell^+\ell^-) = \dfrac{N_{\rm{sig}}}{2 \times f^{+-(00)}\times \varepsilon \times N_{B\overline{B}}}$,
\end{center}
where, $N_{\rm{sig}}$, $f^{+-(00)}$, $\varepsilon$, and $N_{B\overline{B}}$ are the signal yields extracted from the fit, branching fraction of ${\cal B}(\Upsilon(4S) \to B^+B^- (B^0\overline{B^0}))$, signal efficiency corrected for data-MC difference as detailed in section~\ref{systematic}, and number of $B\overline{B}$ pairs derived from a 
data-driven subtraction of the non-resonant contribution from the recorded data, respectively. We use $f^{+-} = (51.4 \pm 0.6)\%$ and $f^{00} = (48.6 \pm 0.6)\%$ for charged and neutral $B$ mesons~\cite{PDG}. The efficiency varies from $6-16\%$ depending on the decay mode and $N_{B\overline{B}}=197 \times 10^{6}$. The branching fractions  for the entire $q^2$ region, excluding the charmonium resonances ($J/\psi$ and $\psi(2S)$) and low $q^2$ region to remove $B \to K^{\ast}\gamma (\to e^+e^-)$ background, are
\begin{align*}
    {\cal B}(B \to K^{\ast}\mu^+\mu^-) &= (1.19 \pm 0.31 ^{+0.08}_{-0.07})\times 10^{-6}, \\
    {\cal B}(B \to K^{\ast}e^+e^-) &= (1.42 \pm 0.48 \pm 0.09) \times 10^{-6}, \\
    {\cal B}(B \to K^{\ast}\ell^+\ell^-) &= (1.25 \pm 0.30 ^{+0.08}_{-0.07}) \times 10^{-6}.
\end{align*}
Here, the first and second uncertainties are statistical and systematic, respectively. The precision of the result is limited by sample size and compatible with world average values~\cite{PDG}.

\section{Acknowledgement}
We thank the SuperKEKB group for the excellent operation of the accelerator; the KEK cryogenics group for the efficient operation of the solenoid; the KEK computer group for on-site computing support.

\end{document}